# Semantic Similarity Computing for Scientific Academic Conferences fused with domain features

Runyu Yu[1], Yawen Li[2]*, Ang Li[1]

[1] （School of Computer Science, Beijing Key Laboratory of Intelligent Telecommunication Software and Multimedia, Beijing University of Posts and Telecommunications, Beijing 100876, China）

[2] （School of Economics and Management, Beijing University of Posts and Telecommunications, Beijing 100876, China）

**Abstract**: Aiming at the problem that the current general-purpose semantic text similarity calculation methods are difficult to use the semantic information of scientific academic conference data, a semantic similarity calculation algorithm for scientific academic conferences by fusion with domain features is proposed. First, the domain feature information of the conference is obtained through entity recognition and keyword extraction, and it is input into the BERT network as a feature and the conference information. The structure of the Siamese network is used to solve the anisotropy problem of BERT. The output of the network is pooled and normalized, and finally the cosine similarity is used to calculate the similarity between the two sessions. Experimental results show that the SBFD algorithm has achieved good results on different data sets, and the Spearman correlation coefficient has a certain improvement compared with the comparison algorithm.

**Key words** science academic conference; deep learning; natural language processing; semantic textual similarity

CLC number TP391

## 1 Introduction

As massive data generated by scientific research-related activities[1][2]. The data of scientific and technological academic conferences include a collection of papers in a certain field. Using natural language processing technology[3] to mine potential information of scientific and technological academic conferences[4], determine the semantic similarity between conferences, and then build knowledge maps and portraits[5], which can help researchers quickly obtain valuable research information[6].

The semantic text similarity calculation mainly includes methods based on strings, based on statistical machine learning, and based on deep learning. At present, the methods based on deep learning are the most widely used and have achieved the best results. However, in the data of scientific and technological academic conferences, the semantic similarity calculation method in the general field cannot mine the potential semantic information and cannot achieve the optimal effect. At the same time, BERT is currently the most outstanding pre-training model in the field of natural language processing, but its performance in semantic text similarity is not well. The domain feature information of the conference is obtained by methods such as identification and keyword extraction, and it is input into the BERT network as a feature and the conference information. The structure of the Siamese network is used to solve the anisotropy problem of BERT, and the output of the network is Pooling and standardization, and using cosine similarity to calculate the similarity between two conferences can effectively improve the computing performance of semantic similarity in scientific and technological academic conferences.

The main contributions of this paper include three aspects:

1) A method for calculating the semantic similarity of scientific and technological academic conferences

Received date: 2022-02-18 ; Revised date:
Funded by: This work is supported by National Key R&D Program of China (2018YFB1402600), the National Natural Science Foundation of China ( 61772083, 61877006, 61802028, 62002027)
Corresponding author: Yawen Li (warmly 0716 @126.com )

that integrates domain features is proposed, which is fine-tuned on the basis of the pre-training model to improve the accuracy of semantic text similarity calculation.

2) The keyword extraction technology is used to obtain the domain information in the conference, and the domain features of the conference are integrated in the sequence input layer to improve the accuracy of semantic text similarity calculation.

3) The Siamese network structure is adopted to solve the problem of BERT's poor performance in similarity calculation, and at the same time improve the calculation speed of the model.

## 2 Related work

Semantic text similarity calculation plays an extremely important role in text classification, text clustering[8][9][10], question answering system [11][12],machine translation [13][14]. The calculation methods of semantic text similarity are mainly based on strings, based on statistics, based on knowledge bases and based on deep learning. Among them, the string-based method is the most intuitive method, which directly compares the original text. The main calculation methods include edit distance[15][16], Jaccard similarity[17], etc. The principle is simple and the implementation is convenient, but only Identify character-level information, generally used for fast text matching. Statistics-based methods mainly include the VSM model and the LDA[18][19] model. The method based on deep learning needs to be carried out on the basis of distributed word vectors. The word vector technology[20] is to map words into vectors that can be recognized by neural networks[21]. The word2vec proposed by Mikolov[22] et al. is the earliest method to generate distributed word vectors, and also provides corresponding tools. Pennington [23] proposed the Glove model. Glove constructed the co-occurrence matrix of words based on the corpus. Using the calculation method of probability theory, combined with the constructed matrix, the final word vector was calculated. Since the construction of the matrix synthesizes the global corpus, Glove considers the global information to a certain extent.

Peters [24] proposed the ELMO model, which first uses the language model to learn the word vectors of good words on a large corpus, at this time, polysemy words cannot be distinguished. Vaswani[25] proposed a Transformer encoder model built on an attention mechanism. Radford[26] proposed the GPT model and introduced the Transformer architecture. The BERT model proposed by Devlin[27] was built on the basis of Transformer, and introduced the idea of mask covering coding and the next sentence prediction method of sentence prediction, and achieved better performance in generating dynamic word vectors. Huang[28] proposed the DSSM method. DSSM is based on the Siamese network architecture. The method model is divided into input layer, representation layer and matching layer. Palangi[29] introduced the LSTM network, which is a special RNN, which can take into account the more distant context information and some sequence information, which improves the effect of the method. [30] used both the CNN model and the LSTM model for the Siamese network architecture, and used the network to calculate the semantic text similarity. Reimers[31] proposed the Sentence-BERT network structure, and the SBERT model only takes 5s to complete, which brings a huge efficiency improvement. Li[32] analyzed from the level of BERT training vector results and found that the word vectors pre-trained by BERT have problems of anisotropy and low-frequency vocabulary sparse, which are better on STS12-16 and SICK-R datasets Performance.

## 3 Calculation method of semantic similarity of scientific and technological academic conferences

### 3.1 Structure of the SBFD method

The results of the BERT model output are corrected smoothly, and considering the characteristics of scientific and technological academic conferences, combined with the research field and direction of the conference, a method for calculating the semantic similarity of scientific and technological academic conferences (SBFD) that integrates the characteristics of the field is proposed. The overall structure of the method is shown in Figure 1. The sequence input layer is calculated by the similarity of scientific and technological academic conferences, the neural network layer is calculated by the similarity of scientific and technological academic conferences, the pooling and normalization layer of similarity calculation of scientific and technological academic conferences, and the similarity calculation of scientific and technological academic conferences are calculated. Consists of four parts.

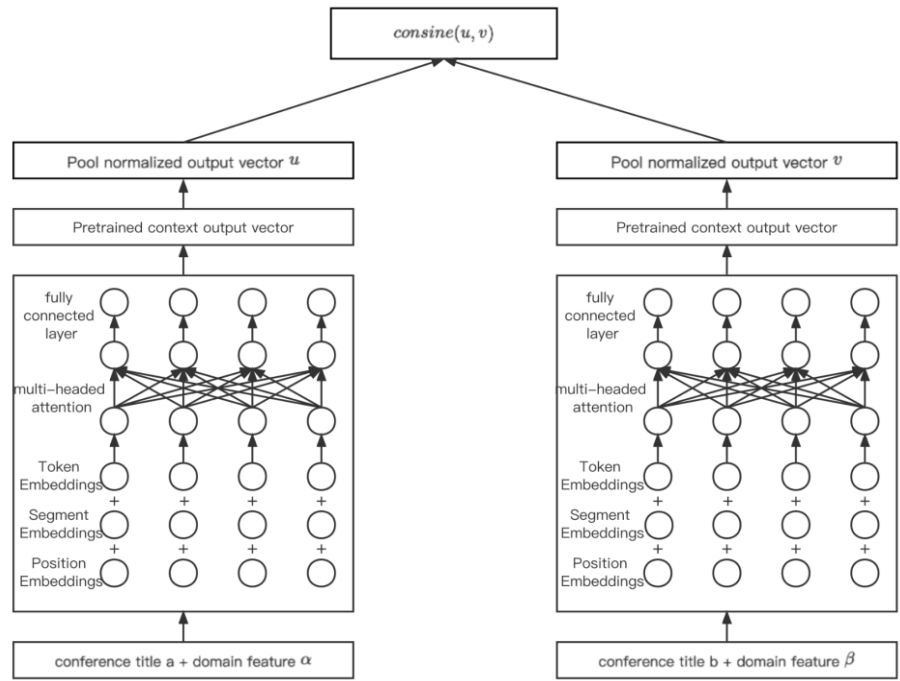

Figure 1 The framework of the SBFD method

## 3.2 Sequence input layer for similarity calculation of scientific and technological academic conferences

The similarity calculation sequence input layer of scientific and technological academic conferences takes the two texts whose similarity needs to be compared as input, and passes them to the Siamese network structure. The standard Siamese network structure is shown in Figure 2, the input is two text sequences to be compared. Where twinning is best manifested is when the network has the same encoder, the part that converts the text into a vector that the neural network can recognize. The network then computes the features of the two texts to complete classification or similarity prediction. The advantage of the Siamese network is that it provides a simple network structure on the basis of ensuring stable training. In the implementation of this method, the sequence is optimized according to the characteristics of the data, because there are two conference research directions in the data with high similarity, but the conference name does not have a particularly high similarity from the text level. In order to reduce this difference. The impact of this paper combines the research domain characteristics of the conference, together as the input of the sequence. Such as conference title 1: International Conference on Computer Design, Conference 2: International Conference on Code Generation and Optimization, all conferences on computer architecture, but the name of the conference does not show the obvious connection between the two conferences, International Conference on Computer Design It is closer to the Asian Computer and Communication Conference, which studies information security. Therefore, the domain information of the conference is also used as the input of the sequence, and is jointly passed to the neural network layer to obtain the vector.

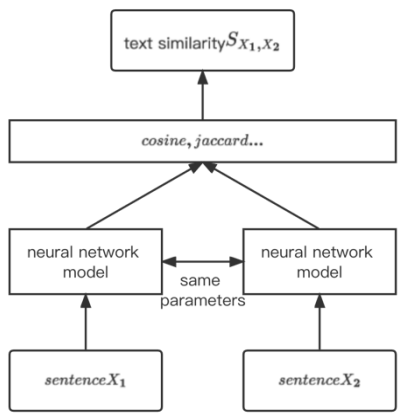

Figure 2 Schematic diagram of the Siamese network structure

### 3.3 Neural network layer for similarity calculation in scientific and technological academic conferences

The network layer based on the Siamese network structure can be implemented by selecting different neural networks, such as the most basic RNN recurrent neural network, or the traditional LSTM, bidirectional LSTM, LSTM + Attention, GRU, etc., which are improved on the basis of RNN. The implementation uses BERT as the network layer. BERT is a pre - training model with bidirectional encoding, and contextual information is also considered during training. Its network architecture is based on the Transformer encoder, using Masked Language Two training tasks, Masked Language Model and Next Sentence Prediction, train network parameters simultaneously. Traditional language models (Language The formula for the acquisition probability of Model, LM) is shown in formula (1).

$$logp(x_{1:T}) = \sum_{t=1}^{T} logp(x_t | c_t) \quad (1)$$

where $X_{1:T} = (x_1,......,x_t)$ represents the token sequence, $c_t = x_{1:t-1}$. When the traditional language model calculates the probability of token occurrence $p(x_{1:T})$, the autoregressive method is used for factorization, for example, the probability of the occurrence of $x_1 x_2 x_3$ is $\cdot \sqrt{p(x_1) * p(x_2 | x_1) * p(x_3 | x_1 x_2)}$

As BERT is a bidirectional language model, it needs to consider the context and context of the word at the same time. To achieve this, the MLM model is used to randomly shield some input tokens, and then the model is trained to correctly predict those shielded tokens. The calculation formula is shown in formula (2).

$$p(\bar{x}\hat{x}) = \sum_{t=1}^{T} m_t p(x_t | c_t) \quad (2)$$

where $\hat{x}$ is with masked sequence of tokens, $\bar{x}$ masked token, $m_t$ For whether the token is masked, there are only two values, 1 is yes, 0 is no.

MLM model is for word-level training. There are many tasks at the sentence level. This requires the language model to understand the relationship between sentences. The next sentence prediction in BERT is to pre-train a binarization. next sentence prediction task. After completing the parameter training of the above two parts, Bert adopts the Transformer structure, in which the core module of the coding unit utilizes the self-attention mechanism. In the BERT model, in order to extend the expressive ability of the model to focus on different positions, the MultiHead mode, is adopted, as shown in formula (3)(4).

$$MutiHead(Q,K,V) = Concat(head_1, \cdots head_n) \quad (3)$$

$$head_i = Attention(Q \cdot W_i^Q, K \cdot W_i^K, V \cdot W_i^V) \quad (4)$$

The output of the attention layer is the high-dimensional vector output by BERT.

### 3.4 Similarity calculation pooling and standard layer in scientific and technological academic conferences

The vectors output by the BERT network may have the problem that the lengths of the vectors after training are different, and it is difficult to calculate the distance between the two tensors of different dimensions. Therefore, the global average pooling is used to extract the semantic representations U and V at the sentence level, and U and V are obtained. Similarity calculations can be performed. Due to the problem of anisotropy in the results after BERT training, the final prediction effect is affected. Therefore, a layer of vector normalization is added here to define a reversible transformation from the observation space of the $u$ latent space. The $z$ generation process of the normalized flow is as follows stated:

$$z \sim p_Z(z), u = f_\phi(z) \quad (5)$$

where $p_Z(z)$ is the prior distribution, which $z->u$ is a reversible transformation.

the probability density function of the observable variable can be expressed as $x$ formula (6).

$$p_U(u) = p_Z\left(f_\phi^{-1}(u)\right)\left|\det\frac{\partial f_\phi^{-1}(u)}{\partial u}\right| \quad (6)$$

The training objective is to maximize the likelihood function of the predetermined BERT sentence vector, as shown in Equation (7).

$$\log p_Z\left(f_\phi^{-1}(u)\right) + \log\left|\det\frac{\partial f_\phi^{-1}(u)}{\partial u}\right| \quad (7)$$

Where D is the dataset, that is, the input sentence set, which $p_Z$ is a standard Gaussian distribution, u is the BERT sentence vector distribution, and det is the determinant of the matrix.

### 3.5 The similarity calculation layer of scientific and technological academic conferences

The cosine similarity puts the text in the vector space, which is more suitable for the data of scientific and technological academic conferences and has strong interpretability. Therefore, it is selected as the similarity calculation layer method of this method to measure the distance between the two final output text vectors.

method steps of semantic similarity of scientific and technological academic conferences integrating domain features

| Method 1: Siamese-BERT semantic similarity calculation method fused with domain features |
|---|
| Input: Document D, contains 2n sentence text sequences $\alpha$ and their semantic features $\beta$, each line has two pairs of text sequences and features, separated by spaces, $\alpha_1$, $\beta_1$, $\alpha_1, \beta_2$ |
| Output: cosine similarity sequence |
| 1: for $(\alpha, \beta) \in D$ do |
| 2: Feature stitching: $(\alpha_1, \beta_1)->e_1, (\alpha_2, \beta_2)->e_2$ |
| 3: Siamese network training: $Bert\_left(e_1)->h_1, Bert\_right(e_2)->h_2$ |
| 4: Pooling Normalization: $mean\ \&\ normalized(h_1)->o_1$, $mean\ \&\ normalized(h_2)->o_2$ |
| 5: Calculate similarity: $cosine\ similarity(o_1, o_2)->v$ |
| 6: Add the result to the list: v add to list |
| 7: return list |

## 4 Experimental results and analysis

### 4.1 Dataset

This experiment is divided into two parts. In order to verify the generalization ability of the method, the method without the fusion of domain features is tested on the public dataset. The datasets used include the widely used STS 12- STS 16 dataset, and SICK-R dataset. The data set of scientific and technological academic conferences is the text content crawled from CNKI and Wanfang. Since the data needs to be manually labeled, 1,000 pieces of data under three disciplines of information technology are selected, of which 8,00 are used for training and 200 are used for training. For testing, due to the limitation of labeling, the overall data volume is limited, the data is cross-validated, and the average result is taken as the final model performance.

### 4.2 Evaluation

There are two main indicators for the calculation of semantic text similarity, namely, the Pearson correlation coefficient and the Spearman correlation coefficient. Studies have shown that the Pearson correlation coefficient is more sensitive to linear relationships, and the intrinsic evaluation of its correlation may produce Misleading, the Pearson correlation coefficient is not the best choice for the task of detecting text similarity, so this experiment mainly uses the Spearman correlation coefficient to evaluate the results.

### 4.3 Experimental parameter settings

The batch size of BERT is set to 32 and the learning rate is set to 1e-5. The pooling method of vector output is global average pooling, the optimizer selects Adam.

**4.4 Experimental results of SBFD method**

The results of text similarity analysis are evaluated using Pearson correlation coefficient and Spearman correlation coefficient, and the effect of different methods is evaluated. In the structure, the effects of GloVe, BERT, SBERT and SBERT methods are compared. Here, SBERT does not input domain information, mainly to verify the ability of the overall network framework. The experimental results are shown in Table 1.

Table 1. Performance of different methods on STS 12-16 and SICK - R

|       | STS12  | STS13  | STS14  | STS15  | STS16  | SICK-R |
|-------|--------|--------|--------|--------|--------|--------|
| GloVe | 0.5525 | 0.6728 | 0.6215 | 0.6746 | 0.6423 | 0.5608 |
| BERT  | 0.4238 | 0.5766 | 0.5825 | 0.6322 | 0.6207 | 0.5889 |
| SBERT | 0.6881 | 0.7276 | 0.7322 | 0.7423 | 0.7133 | 0.7206 |
| SBFD  | 0.6920 | 0.7321 | 0.7426 | 0.7618 | 0.7380 | 0.7354 |

from Table 1, using the siamese network structure to obtain Embedding by sentence semantics, namely SBERT, it can be seen that the effect is significantly improved. SBFD solves the problem of uneven distribution of vector space due to the standardization of the output vector. The best results were obtained. The training was conducted on the data set of the scientific and technological academic conference, and different network structures were selected for the network layer, namely the Spearman coefficients in the case of LSTM, LSTM + Attention, GRU, SBERT, and SBFD. The experimental results are shown in Table 3.

Table 2 Similarity analysis performance of different network structures

|                                  | LSTM   | LSTM+ATT | SBERT  | SBFD_1 | SBFD_2 |
|----------------------------------|--------|----------|--------|--------|--------|
| Information Technology           | 0.6036 | 0.6221   | 0.6682 | 0.7029 | 0.7525 |
| Engineering Technology           | 0.5258 | 0.5882   | 0.6236 | 0.6822 | 0.7057 |
| Agricultural Science and Technology | 0.5620 | 0.6436 | 0.6822 | 0.6918 | 0.7231 |
| average                          | 0.5638 | 0.6178   | 0.6580 | 0.6923 | 0.7271 |

Table 2 that different neural network models have a great influence on the Siamese network in the field of text similarity calculation. SBFD_1 means SBFD method that does not consider domain information, SBFD_2 means SBFD method that includes domain information, SBFD has achieved better results in different network structures. The comparison methods include LSTM, LSTM+ATT. Through LSTM identification, long-distance dependent information and following information may be ignored. Therefore, the attention mechanism is combined on the basis of LSTM. The global semantic information can be considered to improve the effect of similarity calculation. SBERT uses BERT as the neural network layer of the Siamese network to pre-train the model to extract features, and also achieves good results. Finally, compared with the SBFD method, only standardizing the BERT output has achieved better results and solved the problem of vector distribution alienation. The introduction of domain information brings about the improvement of the effect, which verifies the effectiveness of the method in this paper.

**4.5 Influence of experimental parameters on performance**

The batch size is set to different values for experiments to determine its impact.

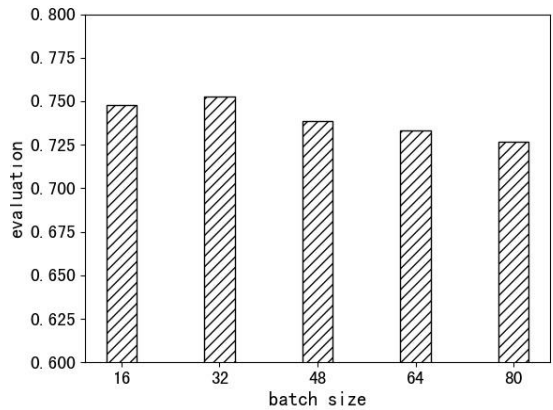

Figure 3 The effect of batch size on the information technology datasets

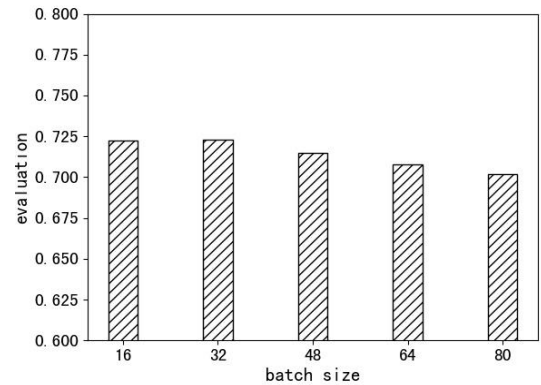

Figure 5 The effect of batch size on the agricultural science and technology datasets

According to Figure 3, it can be seen that in the information technology data set, batch The method achieves the best results when the size is 32 , with the batch With the increase of size, the effect has a certain decline, but the decline is not obvious. Therefore, from the perspective of information technology data sets, batch Size has a certain influence on the effect of the method, but it is not a trend of positive or negative correlation, and the impact is limited.

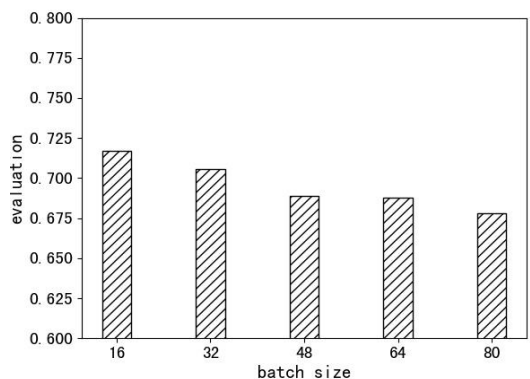

Figure 4 The effect of batch size of engineering science and technology datasets

From Figure 4, in the engineering science and technology dataset, the method achieves the best results when the size is 16, which is different from the information technology data set, but the distribution trend is consistent, that is, batch Size has an impact on performance, but the impact is not obvious. The best effect is achieved at a certain value. As it continues to increase, the effect will decrease compared to 16.

As can be seen from Figure 5, in the agricultural science and technology dataset, batch The best effect is achieved when the size is 32. Combined with the analysis of Figures 3, 4, and 5, it can be concluded that batch size has a certain influence on the recognition effect of the entire network, batch The larger the size, the faster the training speed, but the large batch is will lead to a decrease in the generalization ability of the model. Within a certain range, increase the batch The stability that size contributes to the convergence is as the batch. As the size increases, the performance of the method will decrease, which will affect the accuracy. Select a suitable batch as much as possible within the range allowed by the training speed size, which can improve the performance of the method.

## 5 Conclusion

This paper proposes a method for calculating the semantic similarity of scientific and technological academic conferences that integrates the characteristics of the field. Combined with the characteristics of the scientific and technological academic conferences, they are integrated into the characteristics of the research field, and they are used as vector inputs. The text semantic similarity is calculated based on the Siamese network structure, in which the neural network layer selects the BERT model, making full use of the advantages of the BERT pre-training model to mine the deep semantic information in the text. At the same time, considering the shortcomings of BERT's high computational cost and accuracy in text similarity calculation, based on the Siamese network structure, the SBERT network is constructed, and the trained vectors are standardized to calculate the text similarity. Comparing the SBFD method with Glove, LSTM, GRU and other methods, the experimental results show that SBFD has better performance on academic conference datasets. The results of similarity calculation can be used for the construction of knowledge maps and portraits in academic conferences, helping researchers to quickly obtain the desired scientific research information.

作者简介：

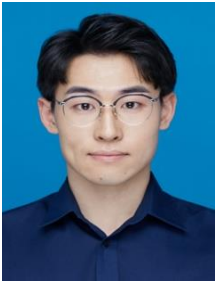

**Runyu Yu**, born in 1997. He is a Master candidate in School of Computer Science(National Pilot School of Software Engineering) of Beijing University of Posts and Telecommunications. The main research directions are deep learning and data mining

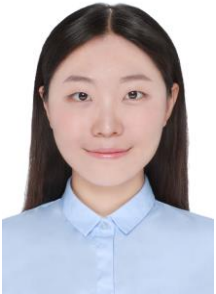

**Yawen Li** (corresponding author), born in 1991. She is now an associate professor of School of economics and management, Beijing University of Posts and telecommunications. The main research directions are enterprise innovation, artificial intelligence, big data, etc. (warmly0716@126.com )

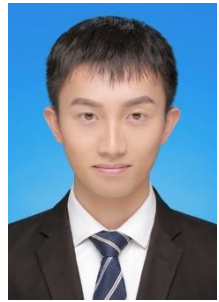

**Ang Li**, born in 1993. He is a Ph.D. candidate in School of Computer Science(National Pilot School of Software Engineering) of Beijing University of Posts and Telecommunications. The main research directions are information retrieval, data mining and machine learning.